\documentstyle[preprint,aps,epsfig]{revtex}
\newcommand \beq{\begin{eqnarray}}
\newcommand \eeq{\end{eqnarray}}

\def\phib{\phi}

\def\d{{\rm d}}

\begin{document}
\preprint{SACLAY-T99/077}
\baselineskip=12pt

\title
 {The transition temperature of the dilute interacting Bose gas for $N$
internal degrees of freedom}
\author{Gordon Baym,$^{a,b,c}$ Jean-Paul Blaizot,$^b$ and Jean Zinn-Justin$^b$
}
\address
      {$^a$University of Illinois at Urbana-Champaign,
      1110 W. Green St., Urbana, IL 61801 \\
$^b$CEA-Saclay,  Service de Physique Th\'eorique,
   91191 Gif-sur-Yvette Cedex, France\\
$^c$Ecole Normale Sup\'erieure, 24 r. Lhomond, 75006 Paris, France\\}


\maketitle

\begin{abstract}
    We calculate explicitly the variation $\delta T_c$ of the Bose-Einstein
condensation temperature $T_c$ induced by weak repulsive two-body interactions
to leading order in the interaction strength.  As shown earlier by general
arguments, $\delta T_c/T_c$ is linear in the dimensionless product $an^{1/3}$
to leading order, where $n$ is the density and $a$ the scattering length.
This result is non-perturbative, and a direct perturbative calculation of the
amplitude is impossible due to infrared divergences familiar from the study of
the superfluid helium lambda transition.  Therefore we introduce here another
standard expansion scheme, generalizing the initial model which depends on one
complex field to one depending on $N$ real fields, and calculating the
temperature shift at leading order for large $N$.  The result is explicit and
finite.  The reliability of the result depends on the relevance of the large
$N$ expansion to the situation $N=2$, which can in principle be checked by
systematic higher order calculations.  The large $N$ result agrees remarkably
well with recent numerical simulations.
\end{abstract}


\section{Introduction}

    The effect of a weak repulsive two-body interaction on the transition
temperature of a dilute gas Bose gas at fixed density has been controversial
for a long time \cite{LY,huang,toyoda,stoof,GCL97,laloe}.  It has recently
been established
theoretically \cite{club} that $T_c$ increases linearly with the strength
of the interaction
parametrized in terms of the scattering length $a$.  However the coefficient
cannot be obtained from perturbation theory.  In Ref.  \cite{club} a simple
self-consistent approximation was used to derive an explicit estimate.

    In this article we first give a more direct derivation of the linear
behavior, using general renormalization group arguments.  Recognizing that the
Hamiltonian of the system under study, which also describes the helium
superfluid transition, is a particular example of the general $N$ vector
model, for $N$ =2, we generalize the problem to arbitrary $N$.  This
generalization makes new tools available; in particular, the coefficient of
$\Delta T_c/T_c$ can be calculated by carrying out an expansion in $1/N$.
Here we calculate explicitly to leading order in $1/N$.  The result happens to
be independent of $N$, for non-trivial reasons.  The calculation involves
subtle technical points which are most easily dealt with by dimensional
regularization.  Surprisingly our result is in remarkable agreement with the
most recent numerical simulations \cite{HM99}.

    This paper is organized as follows:  In Sec.  II we lay out the basics of
the problem.  Then in Sec.  III we present the general $N$ vector model and
analyze the behavior of the temperature shift by renormalization group
arguments.  Finally in Sec.  IV we calculate the leading order contribution.

\section{Classical field theory of the Bose-Einstein condensation}

    We consider a system of identical bosons of mass $m$, at temperature $T$
close to the critical temperature $T_c$.  The effective Hamiltonian of the
system may be written as:
\beq
{\cal H}=\int {\rm d}^3 x
    \left( \frac{\hbar^2}{2m}\nabla \psi^\dagger(x)\cdot \nabla \psi(x) -
    \mu \psi^\dagger(x)\psi(x)
     + \frac{2\pi\hbar^2 a}{m}\,\psi^\dagger(x)\psi^\dagger(x)\psi(x)\psi(x)
     \right),
\eeq
where $\psi^\dagger(x)$ and $\psi(x)$ are the creation and the
annihilation operators of the bosons, $a$ is the scattering length, and $\mu$
is the chemical potential.  Since we are interested in long wavelength
phenomena, we have replaced the two-body potential by a delta function
pseudopotential, with strength proportional to the scattering length $a$.  We
furthermore assume that $a\ll \lambda$, where
\beq
  \lambda=\sqrt{\frac{2\pi \hbar^2}{mk_BT}}
\eeq
is the thermal wavelength.  (In the following we use units $\hbar=k_B=1$,
and write simply $\lambda^2=2\pi/mT$.)

    To compute the effects of the interactions on the transition temperature,
we write the particle number density as a sum of the single particle Green's
function over Matsubara frequencies $\omega_\nu=2\pi i\nu T$:
\beq
n = -T\int\frac{d^3k}{(2\pi)^3} \sum_\nu G(k,\omega_\nu).
\label{matsu0}
\eeq
Above the transition, the single particle  Green's function  obeys the
equation:
\beq
G^{-1}(k,z) = z + \mu - \frac{k^2}{2m} - \Sigma(k,z).
\label{gf}
\eeq
The Bose-Einstein condensation temperature is determined by the point
where $G^{-1}(0,0)=0$, i.e., where $\Sigma(0,0) = \mu$.  At this point,
\beq
G^{-1}(k,z) = z - \frac{k^2}{2m} - \left(\Sigma(k,z)-\Sigma(0,0)\right).
\label{gf1}
\eeq
At $T_c$ the spatial Fourier transform of the two-point correlation
function at zero frequency diverges at zero momentum, and so does the
correlation length.

    In the absence of interactions, $\mu=0$ at the transition, and
\beq
  n=\zeta(3/2)/\lambda_c^3
\eeq
where $\lambda_c^2=2\pi/mT_c^0$, and $T_c^0$ is the condensation
temperature of the ideal gas.

    In the presence of weak interactions, the temperature of the Bose-Einstein
condensation becomes the critical temperature of the interacting model, and is
shifted by the interactions. From the theory of critical phenomena we know that
the variation of the critical temperature in systems with dimension $d$ below
four depends primarily on contributions from the small momenta or large
distance (which we refer to as the infrared, or IR) region.  This property,
which we later verify explicitly for $d=3$, simplifies the problem, since to
leading order the IR properties are only sensitive to the $\omega_\nu=0$
component.

    As shown in Ref.~\cite{club}, in the dilute limit where only the
$\omega_\nu=0$ Matsubara frequency contributes, the shift in the transition
temperature at fixed density, $\Delta T_c = T_c - T_c^0$, can be related to
the change $\Delta n$ in the density at fixed $T_c$ by:
\begin{eqnarray}
  \frac{\Delta T_c}{T_c} =-\frac{2}{3}\frac{\Delta n}{n},
\label{delta}
\end{eqnarray}
where
\beq
\Delta n=\frac{2}{\pi\lambda^2}\int_0^\infty {\rm d}k\,
k^2\left(\frac{1}{k^2+U}-\frac{1}{k^2}\right),
\eeq
and
\beq
U(k)\equiv 2m\left(\Sigma(k)-\Sigma(0)\right).
\eeq

    Once restricted to their zero Matsubara frequency components, the fields
$\psi$ and $\psi^\dagger$ can be considered as classical fields, and the
entire calculation can be cast in terms of classical field theory.  It is then
convenient to rescale the field $\psi$ in order to introduce more conventional
field theory normalizations, and to parametrize it in terms of two real fields
$\phi_1,\phi_2$:  $\psi=\sqrt{mT}(\phi_1+i\phi_2)$.  The partition function
then reads
\beq
 {\cal Z}= \int \left[ \d \phib (x) \right] \exp \left[-{\cal S}(\phib)\right]
 , \label{eONpart}
\eeq
where ${\cal S}(\phib)=H/T$ is given by:
\label{eLGWphi}:
\beq
{\cal S} \left( \phib \right)= \int \left\lbrace{ 1 \over 2} \left[
\partial_{\mu} \phib (x) \right]^2+{1 \over 2}r
\phib^2 (x)+{u \over 4!} \left[ \phib^2(x) \right]^2 \right\rbrace \d^{d}x\,,
\label{eactON}
\eeq
where $r=-2mT\mu$, and:
\beq
u=96 \pi^2\frac{a}{\lambda^2}.
\eeq
In Eq.~(\ref{eactON}) we have kept the dimension $d$ of the spatial
integration arbitrary in order to use dimensional regularization later.
The single particle Green's function $G(p)$ is related to the inverse
two-point function $\Gamma^{(2)}(p)$ of the classical field theory by
\beq
G(p)=\frac{2mT}{ \Gamma^{(2)}(p)}.
\eeq

    The model described by the Euclidean action (\ref{eactON}) reduces to the
ordinary $O(2)$ symmetric $\phib^4$ field theory, which indeed describes the
universal properties of the superfluid helium transition.  As it stands this
field theory suffers from UV divergences.  These are absent in the
original theory, the higher frequency modes providing a large momentum
cutoff $\sim \sqrt{m T}\sim 1/\lambda$.  This cutoff may be restored when
needed, e.g., by replacing the propagator by the regularized propagator:
\beq
{2mT \over  k^2}\rightarrow {1\over {\rm e}^{k^2/2mT}-1}\,.
\eeq
In fact, as we show later, since the shift of the critical temperature is
dominated by long distance properties it is independent of the precise
cutoff procedure.

    A second effect of the non-zero frequency modes is to renormalize the
effective coefficients of the Euclidean action.  This problem can be explored
by writing the functional integral representation of the complete quantum
theory and integrating over the non-zero modes perturbatively.  The
corrections generated are of higher order in $a$ and can thus be neglected.

    Because the interactions are weak, one may imagine calculating the change
in the transition temperature by perturbation theory.  However the
perturbative expansion for a critical theory does not exist for any fixed
dimension $d<4$; IR divergences prevent a complete calculation.  If one
introduces an infrared cutoff $k_c$ to regulate the momentum integrals, one
finds that perturbation theory breaks down when $k_c \sim a/\lambda^2$, all
terms being then of the same order of magnitude.  To discuss this problem in
more detail we now generalize the model to $N$ component fields with an $O(N)$
symmetric Hamiltonian.

\section{The $N$-vector model. Renormalization group}

    We consider the $O(N)$ symmetric generalization of the model
corresponding to the Euclidean action (\ref{eactON}).  The field $\phib(x)$
then has $N$ real components, and, e.g.,
\beq
\phib^2=\sum_{i=1}^N \phi_i^2\ .
\eeq
The action ${\cal S}(\phib)$ is still given by Eq.~(\ref{eactON}), now with
an $O(N)$ symmetry. The advantage of this generalization is that it provides
us with a tool, the large $N$ expansion, which allows us to calculate at
the critical point \cite{rlargeN,rDoGriv,rZJTai}.

    The goal is to obtain the leading order non-trivial contribution at
criticality (in the massless theory) to
\beq
n=2mT \sum_{i=1}^N  \left<\phi_i^2 \right> \equiv {2mT \,
N\,}\rho,
\eeq
with
\beq
\rho= \int\frac{\d^d k}{(2\pi)^d}\,\frac{1}{\Gamma^{(2)}(k)}\label{eONrho}.
\eeq
Here $\delta_{ij}/{\Gamma^{(2)}(k)}$ is the connected two-point correlation
function.

    We now recover, by a simple renormalization group analysis, the result of
\cite{club} that the change in $\rho$ due to the interaction is linear in the
coupling constant.  We introduce a large momentum cutoff $\Lambda
\propto\sqrt{mT}\sim 1/\lambda$, and the dimensionless coupling constant $g$
\beq\label{coupling}
g=\Lambda^{d-4}u\propto \left(a\over \lambda\right)^{d-2}\,.
\eeq
At $T_c$ the inverse two-point function in momentum space satisfies the
renormalization group equation \cite{zinn}
\beq
\left(\Lambda{\partial\over\partial \Lambda}+\beta(g){\partial \over\partial
g}-\eta(g)\right)\Gamma^{(2)}(p,\Lambda,g)=0\,.
\eeq
This equation with dimensional analysis implies that the two-point
function has the general form
\beq
\Gamma^{(2)}(p,\Lambda,g)=p^2 Z(g)F\bigl(p/\Lambda(g)\bigr);
\label{f}
\eeq
on dimensional grounds $\Lambda(g)$ is proportional to $\Lambda$, with
\begin{mathletters}
\beq
\beta(g){\partial \ln Z(g)\over \partial g}&=& \eta(g),
\label{beta}
\\
\beta(g){\partial \ln \Lambda (g)\over \partial g}&=& -1 \,.
\label{lam}
\eeq
\end{mathletters}
Since $\beta(g)=-(4-d) g +(N+8)g^2/48\pi^2+{\cal O}(g^3)$, $\beta(g)$ is
of order $g$ for for small $g$ in $d<4$; similarly
$\eta(g)=(N+2)g^2/(72(8\pi^2)^2)+{\cal O}(g^3)$.  Therefore
\beq
 Z(g)=\exp\int_0^g{\eta(g') \over \beta(g')}\d g'=1+{\cal O}(g^2);
\eeq
to leading order $Z(g)$ =1.  The function $\Lambda(g)$ is then obtained by
integrating Eq.~(\ref{lam}),
\beq
\Lambda(g)=g^{1/(4-d)}\Lambda\exp\left[-\int_0^g\d g'\left({1\over
\beta(g')} +{1\over (4-d)g'}\right)\right].
\eeq
The scale $\Lambda(g)$ plays a specific role in the analysis as the
crossover separating a universal long-distance regime, where
\beq\label{IRscaling}
\Gamma^{(2)}(p)\propto p^{2-\eta} \qquad p\ll \Lambda(g) ,
\eeq
governed by the non-trivial zero, $g^*$, of the $\beta$-function, from a
universal short distance regime governed by the Gaussian fixed point, $g=0$,
where
\beq
\Gamma^{(2)}(p)\propto p^2 \qquad \Lambda(g) \ll p\ll \Lambda\,.
\eeq
However such a regime exists only if $\Lambda(g)\ll \Lambda$, i.e., if
there is an intermediate scale between the IR and the microscopic scales;
otherwise only the IR behavior can be observed.  In a generic situation $g$ is
of order unity, and thus $\Lambda(g)$ is of order $\Lambda$, and the universal
large momentum region is absent.  Instead $\Lambda(g)\ll \Lambda$ implies
\beq
g^{1/(4-d)}\exp\left[-\int_0^g\d g'\left({1\over
\beta(g')} +{1\over (4-d)g'}\right)\right]\ll 1\,.
\eeq
Since $g$ (equal to $a/\lambda$, see Eq.~(\ref{coupling})) is $\ll 1$, this
condition is satisfied in the present situation.

    We now show that with this condition, $\Delta T_c\propto \Lambda(g)$.
First, from the $g=0$ limit we infer that $F(\infty)=1$ (see Eq.~(\ref{f}).)
In three dimensions the function $F(p)$ behaves for large $p$ as
\beq
 F(p)=1+{\cal O}(\ln p/p^2),
\eeq
as can be seen directly from 3-d perturbation theory.
Therefore the first correction to the density (\ref{eONrho}) is convergent at
large momentum and independent of the cutoff procedure,
\beq
\delta\rho= \int\frac{\d^3 p}{(2\pi)^3} {1\over p^2}\left(
{1\over F(p/\Lambda(g))} -{ 1}\right).
\eeq
Similarly the IR scaling result (Eq.~\ref{IRscaling}) implies that this
integral is IR convergent.  Setting $p=\Lambda(g)k$, we then find the general
form
\beq\label{deltarho}
\delta\rho=\Lambda(g)\int \frac{\d^3 k}{(2\pi)^3} {1\over
  k^2}\left( {1\over F(k)} -{ 1}\right);
\eeq
the $g$ dependence is entirely contained in $\Lambda(g)$.  For $g$ small
we conclude
\beq
\frac{\delta\rho}{\rho}\propto g \propto an^{1/3}\,,
\eeq
in agreement with \cite{club}.  It is important to note that both the
perturbative large momentum region and the non-perturbative IR region
contribute to the integral in Eq.~(\ref{deltarho}).  Therefore we cannot
evaluate it from a perturbative calculation of the function $F(p)$.  However,
we now show, we can calculate $\delta\rho$ exactly in an $1/N$ expansion.

\section{The large $N$ expansion at order $1/N$}

    Critical phenomena can be studied in any fixed dimension by the now
standard technique of large $N$ expansion, where the large $N$ limit is taken
at $Nu$ fixed.  To leading order the critical two-point function has simply
its free field form.  However a non-trivial correction is generated at order
$1/N$; one finds the inverse two-point function \cite{aharo,zinn},
\beq
\Gamma^{(2)}(p)= p^2 +{2\over N }\int
 \frac{\d^{d}q}{(2\pi)^{d}}{1 \over(6/Nu)+B(q)}\left({1 \over
(p+q)^2} -{1 \over q^2}\right) +{\cal O}\left({1 \over N^2}\right),
\label{eONpropi}
\eeq
where $B(q)$ is the one-loop contribution to the four-point function
\beq
B(q)=\int \frac{\d^{d}k}{(2\pi)^{d}}\frac{1}{k^2(k+q)^2}\, \cdot
\eeq
Note that in the large $N$ limit, the chemical potential $\mu$ is
proportional to $1/N$.  Using the large $N$ value of $g^*$, and setting
$\varepsilon=4-d$, we can write $B(q)$ for $q\to 0$:
\beq
B(q)=b(\varepsilon)q^{-\varepsilon} -{6\over N g^*}\Lambda^{-\varepsilon}
+{\cal O}(1/\Lambda^2),
\eeq
with
\beq
b(\varepsilon)=-{2
\over\sin(\pi d/2)}  {\Gamma^2 (d/ 2) \over \Gamma (d-1)} N_d\,,
\label{econstb}
\eeq
where $N_d$ is the usual loop factor
\beq
N_d={2 \over(4\pi)^{d/2} \Gamma(d/2)}\,,\quad N_3={1\over
  2\pi^2}\,.
\eeq
For $d=3$, $b(1)=1/ 8$.

    In the large $N$ limit the $\beta$-function takes the simple form
$\beta(g)=-\varepsilon g (1-g/g^*)$.  Therefore the leading cutoff-dependent
correction to $B(q)$ combines with $6/Nu$ to yield
$(6/N)\Lambda^{-\varepsilon}(g)$, as expected from renormalization
group arguments. This cutoff dependent correction can be neglected for small
$g$.

    We evaluate
\beq
\delta\rho=-{2\over N }\int\frac{\d^d p}{(2\pi)^{2d}}{1 \over p^4}
{\d^{d}q \over(6/Nu)+b(\varepsilon)q^{-\varepsilon}}\left({1 \over
(p+q)^2} -{1 \over q^2}\right)
\label{ro}
\eeq
by keeping the dimension $d$ generic and using dimensional regularization
\cite{dimreg}.  The integral over $p$ is
\beq
 \int\frac{\d^d p}{(2\pi)^d}{1 \over p^4}{1\over(p+q)^2}&=&
{1\over(4\pi)^{d/2}}{\Gamma(3-d/2)\Gamma(d/2-1)\Gamma(d/2-2)
\over \Gamma(d-3)}q^{d-6}\,. \nonumber\\
&=&{1\over(4\pi)^{d/2}}{\Gamma(d/2-1)\over\Gamma(d-3)}{\pi\over\sin \pi d/2}\,
q^{d-6}\,.
\eeq
Note that the singularity at $d=3$, which would apparently entail the
vanishing of the integral, is cancelled in the subsequent $q$ integral, which
reduces to
\beq
\int\frac{\d^{d}q}{(2\pi)^{d}}\, {q^{d-6}
\over(6/Nu)+b(\varepsilon)q^{-\varepsilon}} =
{N_d \over 4-d} \quad {\pi\over\sin\left(\pi(d-2)/(4-d)\right)} \quad
b^{(2d-6)/(4-d)}\left(6\over Nu\right)^{(2-d)/(4-d)}\,,
\eeq
In the $d=3$ limit the two integrations in Eq.~(\ref{ro}) yield
$(1/32\pi^2)(Nu/6)$. As expected, $\delta\rho\propto u$:
\beq
\delta\rho=-Ku \,,\quad K={1\over96\pi^2} \,,
\eeq
or in terms of the original parameters,
\beq\label{delrhoa}
\delta\rho=-\frac{a}{\lambda^2}.
\eeq

    It is instructive to repeat the calculation directly at $d=3$.  To do
this, we first eliminate the constant term (which does not contribute), and
write $U(k)$, which equals $\Gamma^{(2)}(k)-k^2$, as
\beq\label{sigsumnew}
U(k)=-2N\left(\frac{u}{6}\right)^2\int \frac{{\rm
d}^3q}{(2\pi)^3}\frac{B(q)}{1+(Nu/6)B(q)}\left(\frac{1}{(k-q)^2}-\frac{1}{q^2}
\right)
\eeq
From Eq.~(\ref{sigsumnew}) we find
\beq
U(k)= -32\pi^2Nk\frac{a^2}{\lambda^4}\int_0^\infty
         \frac{\d x}{kx + 2\pi^2aN/\lambda^2}
 \left(\frac{x}{2}\ln\left|\frac{1+x}{1-x}\right| -1\right).
\eeq
Calculating the change in the density in leading order in $U\sim 1/N$, one
gets:
\beq
\delta\rho= \frac{8a}{\pi^2\lambda^2}\int_0^\infty \frac{{\rm
d}k}{k}\int_0^\infty
         \frac{\d x}{1+\tau xk}
          \left(\frac{x}{2}\ln\left|\frac{1+x}{1-x}\right|
     -1\right),
\eeq
where $\tau = \lambda^2/(2\pi^2 a N)$.   The trick now is to exchange the
orders of the $k$ and $x$ integrals.  In three dimensions, however, the
integrals are not absolutely convergent, preventing this interchange.  Thus we
 introduce a regularization, with by inserting a factor
$k^\epsilon$ in the $k$ integral, and take the limit $\epsilon\to
0^+$.  With this factor we may exchange the orders of integration.
For small $\epsilon$ the $k$ integral becomes
\beq
\int_0^\infty \d k \frac{k^{\epsilon - 1}}{1+\tau x k} =
\frac{1}{\epsilon (\tau x)^\epsilon}.
\eeq
The factor $\tau^\epsilon$ goes to unity, and the remaining $x$ integral
becomes
\beq
  \int \d x\, x^{-\epsilon}
          \left(\frac{x}{2}\ln\left|\frac{1+x}{1-x}\right| -1\right).
\eeq
For $\epsilon = 0$ this integral vanishes identically.  Thus we may replace
$x^{-\epsilon}$ by $x^{-\epsilon}-1$ which goes to $-\epsilon \ln x$ as
$\epsilon \to 0$.  The remaining integral is
\beq
  \int \d x\, \ln x
      \left(\frac{x}{2}\ln\left|\frac{1+x}{1-x}\right| -1\right) = -\frac
{\pi^2}{8}.
\eeq
The factors of $\epsilon$ cancel out, and we find, as before
(Eq.~(\ref{delrhoa})),
$\delta\rho=-a/\lambda^2$.

    Using this result in Eq.~(\ref{delta}), we finally obtain the change in
the transition temperature:
\beq
\frac{\Delta T_c}{T_c} &=&
  \frac{8\pi}{3\zeta(3/2)}\frac{a}{\lambda}
 \nonumber \\ &=&  \frac{8\pi}{3\zeta(3/2)^{4/3}} a n^{1/3}
 = 2.33\, a n^{1/3},
\label{result}
\eeq

    Note that although the final result does not depend on $N$ and therefore
replacing $N$ by two is easy, the result is only valid for $N$ large.  The
result (\ref{result}) is in remarkable agreement with the ($N=2$) value
$\Delta T_c/T_c^0 \approx (2.2\pm 0.2) an^{1/3}$ in the recent numerical
simulations of Holzmann and Krauth \cite{HM99}.

\section{Conclusion}

    In this paper we have shown that the properties of the weakly
interacting Bose gas remain dominated by the UV fixed point of the
renormalization
group equations up to very large length scales; this is why we can still
refer to the
Bose-Einstein condensation when discussing the phase transition of the
dilute interacting
Bose gas.  Renormalization group arguments also enabled us to confirm
directly that the
shift of the transition temperature at fixed density is proportional to the
dimensionless
combination
$an^{1/3}$ for weak interactions.  This result is non-perturbative, and the
proportionality
coefficient cannot be obtained from perturbation theory.  We have therefore
introduced a non-perturbative method, the large $N$ expansion, which allows a
systematic calculation of this coefficient as a power series in $1/N$, where
eventually one has to set $N=2$.  Finally we have calculated explicitly the
leading order contribution.  The first correction is formally of order $1/N$
multiplied by a function of the product $aN$ which is fixed in the large $N$
limit.  Because the final result in three dimensions is linear in $a$, the
$1/N$ factor somewhat surprisingly cancels, and the result is independent of
$N$.  The value found is in remarkable agreement with the most recent
numerical estimates.  Whether this agreement is just a coincidence or reflects
the smallness of the next order (of order $1/N$) correction can only be
checked by an explicit calculation.

\medskip
{\bf Acknowledgments.}
    This work was facilitated by the Cooperative Agreement between the
University of Illinois at Urbana-Champaign and the Centre National de la
Recherche Scientifique, and also supported in part by National Science
Foundation Grant PHY98-00978.


\begin{references}

    \bibitem{LY} T.D.  Lee and C.N.  Yang, Phys.  Rev.  {\bf 112}, 1419
(1957).

    \bibitem{huang} K. Huang, in {\it Stud.  Stat.  Mech}, {\bf II}, J. de
Boer and G.E.  Uhlenbeck, eds.  (North Holland Publ., Amsterdam, 1964), 1.

    \bibitem{toyoda} T.~Toyoda, Ann.  Phys.  (NY) {\bf 141}, 154 (1982).

    \bibitem{stoof} H.T.C.  Stoof, Phys.  Rev.  {\bf A45}, 8398 (1992); M.
Bijlsma and H.T.C.  Stoof, Phys.  Rev.  {\bf A54}, 5085, 1996.

    \bibitem{GCL97} P. Gr\"uter, D. Ceperley, and F. Lalo\"e, Phys.  Rev.
Lett.{\bf 79}, 3549, 1997.

    \bibitem{laloe} M. Holzmann, P. Gr\"uter, and F. Lalo\"e,
cond-mat/9809356, {\it Euro.  Phys.  J.} B (in press).

    \bibitem{club} G. Baym, J.-P.  Blaizot, M. Holzmann, F. Lalo\"e, and D.
Vautherin.  Submitted to {\it Phys.  Rev.  Letters}, cond-mat/9905430.

    \bibitem{HM99} M. Holzmann and W. Krauth, cond-mat/9905198.

    \bibitem{zinn} J.~Zinn-Justin, {\it Quantum Field Theory and Critical
Phenomena}, Clarendon Press (Oxford 1989, third ed.~1996).


    \bibitem{rlargeN} H.E.  Stanley, {\it Phys.  Rev.} 176 (1968) 718; R. Abe,
Prog.  Theor.  Phys. 48 (1972) 1414; {\it ibidem}\/ 49 (1973) 113; S.K.  Ma,
{\it Phys.  Rev.  Lett.} 29 (1972) 1311, {\it Phys.  Rev.} A7 (1973) 2172; M.
Suzuki, {\it Phys.  Lett.} 42A (1972) 5, {\it Prog.  Theor.  Phys.} 49 (1973)
424; K.G.  Wilson, {\it Phys.  Rev.} D7 (1973) 2911.

    \bibitem {rDoGriv} See also the contributions of S.K.  Ma and E. Br\'ezin,
J.C.  Le Guillou and J. Zinn-Justin to {\it Phase Transitions and Critical
Phenomena} vol. 6, C. Domb and M.S. Green eds.  (Academic Press, London 1976).

    \bibitem{rZJTai} For a recent review see J.~Zinn-Justin, {\it Vector
models in the large $N$ limit:  a few applications}, lecture notes of the
11$^{\rm th}$ Taiwan Spring School, Taipei 1997, Saclay preprint SPhT/97-018,
hep-th/9810198.

    \bibitem{aharo} R.A.  Ferrel and D.J.  Scalapino, {\it Phys.  Rev.  Lett.}
29 (1972) 413; A. Aharony, {\it Phys.  Rev.} B10 (1974) 2834.

    \bibitem{dimreg} J. Ashmore, {\it Lett.  Nuovo Cimento} 4 (1972) 289;
G.~'t~Hooft and M.~Veltman, {\it Nucl.  Phys.} B44 (1972) 189; C.G.~Bollini
and J.J.~Giambiaggi, {\it Phys.  Lett.} 40B (1972) 566, {\it Nuovo Cimento}
12B (1972).
\end{references}
\end{document}